\documentclass[review]{elsarticle}

\usepackage{lineno,hyperref}
\modulolinenumbers[5]
\usepackage{multirow}
\usepackage[utf8]{inputenc}

\journal{Physics Letters B}

\newcommand{\invnb}{{nb}$^{-1}$}
\newcommand{\invpb}{{pb}$^{-1}$}
\newcommand{\gev}{GeV/$c$}
\newcommand{\gevtwo}{GeV/$c^{2}$}

\newcommand{\jpsi}{\ensuremath{J/\psi}}

\newcommand{\pAu}{$p$+Au}
\newcommand{\dAu}{$d$+Au}	
\newcommand{\AuAu}{Au+Au}

\newcommand{\pp}{$p$+$p$}

\newcommand{\sqrtsNN}{\ensuremath{\sqrt{s_{_\mathrm {NN}}}}}
\newcommand{\sqrts}{\ensuremath{\sqrt{s}}}

\newcommand{\pT}{\ensuremath{p_\mathrm{T}}}

\newcommand{\nsigmapi}{\ensuremath{\mathrm{n}\sigma_{\pi}}}
\newcommand{\deltaz}{\ensuremath{\Delta z}}
\newcommand{\deltaznorm}{\ensuremath{\Delta z/\sigma_{\Delta z}}}
\newcommand{\deltay}{\ensuremath{\Delta y}}

\newcommand{\deltayqnorm}{\ensuremath{\Delta y\times q/\sigma_{\Delta y\times q}}}
\newcommand{\timeofflight}{\ensuremath{t_{\rm{tof}}}}	
\newcommand{\deltatof}{\ensuremath{\Delta \timeofflight}}

\newcommand{\RAA}{\ensuremath{R_{\rm{AA}}}}
\newcommand{\rpa}{\ensuremath{R_{p\rm{Au}}}}
\newcommand{\rda}{\ensuremath{R_{d\rm{Au}}}}

\newcommand{\ncoll}{\ensuremath{N_{\rm{coll}}}}
\newcommand{\npart}{\ensuremath{N_{\rm{part}}}}

\begin{document}

\begin{frontmatter}

\title{Measurement of cold nuclear matter effects for inclusive \jpsi\ in \pAu\ collisions at \sqrtsNN\ = 200 GeV}

\author{
M.~S.~Abdallah$^{5}$,
B.~E.~Aboona$^{55}$,
J.~Adam$^{6}$,
L.~Adamczyk$^{2}$,
J.~R.~Adams$^{39}$,
J.~K.~Adkins$^{30}$,
G.~Agakishiev$^{28}$,
I.~Aggarwal$^{41}$,
M.~M.~Aggarwal$^{41}$,
Z.~Ahammed$^{60}$
% ...
}

\address{\rm{(STAR Collaboration)}}

% ...
\address{$^{2}$AGH University of Science and Technology, FPACS, Cracow 30-059, Poland}
% ...
\address{$^{5}$American University of Cairo, New Cairo 11835, New Cairo, Egypt}
\address{$^{6}$Brookhaven National Laboratory, Upton, New York 11973}
% ...
\address{$^{28}$Joint Institute for Nuclear Research, Dubna 141 980, Russia}
% ...
\address{$^{30}$University of Kentucky, Lexington, Kentucky 40506-0055}
% ...
\address{$^{39}$Ohio State University, Columbus, Ohio 43210}
% ...
\address{$^{41}$Panjab University, Chandigarh 160014, India}
% ...
\address{$^{55}$Texas A\&M University, College Station, Texas 77843}
% ...
\address{$^{60}$Variable Energy Cyclotron Centre, Kolkata 700064, India}
% ...


\begin{abstract}
Measurement by the STAR experiment at RHIC of the cold nuclear matter (CNM) effects experienced by inclusive \jpsi\ at mid-rapidity in 0-100\% \pAu\ collisions at \sqrtsNN\ = 200 GeV is presented. Such effects are quantified utilizing the nuclear modification factor, \rpa, obtained by taking a ratio of \jpsi\ yield in \pAu\ collisions to that in \pp\ collisions scaled by the number of binary nucleon-nucleon collisions. The differential \jpsi\ yield in both \pp\ and \pAu\ collisions is measured through the dimuon decay channel, taking advantage of the trigger capability provided by the Muon Telescope Detector in the RHIC 2015 run. Consequently, the \jpsi\ \rpa\ is derived within the transverse momentum (\pT) range of 0 to 10 \gev. A suppression of approximately 30\% is observed for $\pT<2$ \gev, while \jpsi\ \rpa\ becomes compatible with unity for \pT\ greater than 3 \gev, indicating the \jpsi\ yield is minimally affected by the CNM effects at high \pT. Comparison to a similar measurement from 0-20\% central \AuAu\ collisions reveals that the observed strong \jpsi\ suppression above 3 \gev\ is mostly due to the hot medium effects, providing strong evidence for the formation of the quark-gluon plasma in these collisions. Several model calculations show qualitative agreement with the measured \jpsi\ \rpa, while their agreement with the \jpsi\ yields in \pp\ and \pAu\ collisions is worse. 
\end{abstract}

\begin{keyword}
RHIC, cold nuclear matter effects, \jpsi\ suppression
\end{keyword}

\end{frontmatter}


\section{Introduction}
In ultra-relativistic heavy-ion collisions, a new state of matter, referred to as the Quark-Gluon Plasma (QGP), is created, in which the deconfined quarks and gluons are the relevant degrees of freedom. With the start of data-taking at the Relativistic Heavy Ion Collider (RHIC) in 2000 and the Large Hadron Collider (LHC) in 2010, tremendous progress has been made in understanding the properties of the QGP. Among various probes used to study the QGP, quarkonia play a unique role as they are expected to be dissociated by surrounding partons, i.e. gluons and quarks, if the medium temperature exceeds the melting temperature of the quarkonium states \cite{Matsui:1986dk,Andronic:2015wma}. Therefore, observations of quarkonium suppression in heavy-ion collisions have been considered strong evidence for QGP formation and important probes of the medium temperature, a fundamental property of the QGP. However, there are other effects that could modify the observed quarkonium yield in heavy-ion collisions, including the main contributions which are recombination and Cold Nuclear Matter (CNM) effects. The former refers to the quarkonium production mechanism arising from combination of deconfined heavy quarks and anti-heavy quarks in the medium, while the latter is due to the participation of nuclei in the collisions, but not as a result of the creation of the QGP.

As the most abundantly produced quarkonium state that is experimentally accessible, the \jpsi\ meson suppression in heavy-ion collisions has been extensively measured at Super Proton Synchrotron, RHIC and the LHC \cite{NA50:2000brc, Adamczyk:2012ey, Adamczyk:2013tvk, Adamczyk:2016srz, Adam:2019rbk, Adare:2006ns, Adare:2011yf, Adam:2015rba, Adam:2015isa, Adam:2016rdg, Khachatryan:2016ypw, Sirunyan:2017isk,Aaboud:2018quy}. At high transverse momentum ($\pT>5$ \gev), \jpsi\ mesons are strongly suppressed in central heavy-ion collisions, which is mainly attributed to the dissociation effect. In order to substantiate this conclusion, precise measurements of the CNM effects are needed to understand their potential contribution to the high-\pT\ \jpsi\ suppression observed in heavy-ion collisions. The CNM effects have been measured through collisions of a nucleus with a proton/deuteron at RHIC and with a proton at the LHC, in which the QGP is not expected to be produced \cite{Adamczyk:2016dhc, Acharya:2019zjt, Adare:2012qf, Acharya:2018yud, Aaboud:2017cif, Sirunyan:2017mzd,ALICE:2018mml}; or even if produced in such small system collisions, it is not expected to have a substantial effect. A general feature in these measurements is that a sizable suppression of the \jpsi\ yield, relative to that in \pp\ collisions, is seen at low \pT, which gradually diminishes with increasing \pT. A hint of a mild enhancement is seen above 10 \gev\ at the LHC energies \cite{Aaboud:2017cif, Sirunyan:2017mzd}. Different physics mechanisms could contribute to the experimental observation. The nuclear parton distribution function (nPDF) is believed to be modified compared to the PDF of a free nucleon, e.g. a suppression (shadowing) at small Bjorken $x$ and an enhancement (anti-shadowing) at intermediate $x$ \cite{Kovarik:2015cma, Eskola:2016oht}. Such a small-$x$ effect, as well as the higher twist contribution \cite{Mueller:1985wy}, can be accounted for alternatively within the framework of the Color Glass Condensate (CGC) effective theory \cite{Gelis:2010nm}. Before forming a bound state, the color-octet $c\bar{c}$ pairs could undergo energy loss within the cold nuclear matter of a nucleus \cite{Arleo:2013zua}. After being formed, the bound-state \jpsi\ meson can break up through interactions with the nucleons in the nucleus \cite{Vogt:1999cu} or due to interactions with co-moving particles produced in the same collisions \cite{Ferreiro:2014bia}. It could also be possible that a small droplet of the QGP is formed in $p/d$+A collisions, leading to \jpsi\ dissociation \cite{Du:2015wha, Du:2018wsj}. 

In this letter, the first measurement of the CNM effects experienced by inclusive \jpsi\ at mid-rapidity in \pAu\ collisions at \sqrtsNN\ = 200 GeV with the Solenoidal Tracker At RHIC (STAR) experiment \cite{Ackermann:2002ad} is presented. They are quantified using the nuclear modification factor (\rpa):
\begin{equation}
\rpa=\frac{1}{\langle T_{\mathrm{AA}}\rangle}\times\frac{(\frac{d^{2}N_{\jpsi}}{d\pT dy})_{p+\rm{Au}}}{(\frac{d^{2}\sigma_{\jpsi}}{d\pT dy})_{p+p}}
\label{eq:rpa}
\end{equation}
where \mbox{$(\frac{d^{2}\sigma_{\jpsi}}{d\pT dy})_{p+p}$} is the \jpsi\ cross section in \pp\ collisions and \mbox{$(\frac{d^{2}N_{\jpsi}}{d\pT dy})_{p+\rm{Au}}$} is the invariant yield per inelastic \pAu\ collision. The nuclear thickness function $\langle T_{\mathrm{AA}}\rangle = \langle\ncoll\rangle/\sigma_{\mathrm{NN}}^{\rm{inel}}$ is calculated using a Glauber model \cite{Miller:2007ri}, where $\sigma_{\mathrm{NN}}^{\rm{inel}}=42$ mb \cite{STAR:2003fka} is the inelastic cross section of nucleon-nucleon collisions at 200 GeV, and $\langle\ncoll\rangle=4.7\pm0.3$ is the average number of binary nucleon-nucleon collisions for 0-100\% \pAu\ collisions \cite{Acharya:2019zjt}. The inclusive \jpsi\ sample used in this analysis includes both directly produced \jpsi\ as well as those from decays of excited charmonium states  (approximately 40\% \cite{Adare:2011vq}) and b-hadrons. Compared to previous measurements of \rda\ at RHIC \cite{Adamczyk:2016dhc,Adare:2012qf}, the new \rpa\ measurement has better precision over the entire kinematic range, especially for \pT\ larger than 3 \gev. This precision is partially achieved as the reference \jpsi\ cross section from \pp\ collisions was recorded the same year with the same trigger set-up and detector configuration as for the \pAu\ collisions, which allows for the partial cancelation of systematic uncertainties

\section{Experiment, data set}
Both the \pp\ and \pAu\ data samples used in this analysis were taken in 2015 by the STAR experiment at RHIC with the ``dimuon" trigger. This trigger is dedicated to quarkonium measurements, and requires a coincidence signal in the east and west Vertex Position Detectors (VPD) \cite{Llope:2014nva} as well as two muon candidates in the Muon Telescope Detector (MTD) \cite{Ruan:2009ug}. The VPD, made of plastic scintillators, covers full azimuth within the pseudorapidity ($\eta$) range of $4.24<|\eta|<5.1$, while the azimuthal coverage of the MTD, consisting of multigap resistive plate chambers, is about 45\% within $|\eta|<0.5$. A hit in the MTD is classified as a muon candidate online if the difference between its arrival time measured by the MTD and the collision start time measured by the VPD falls within a pre-defined window, which is chosen to maximize the trigger efficiency (close to 100\%) while maintaining a reasonable trigger rate. The sampled luminosities online are 122 \invpb\ and 410 \invnb\ for the \pp\ and \pAu\ data sets, respectively. 

The Time Projection Chamber (TPC) \cite{Anderson:2003ur}, encompassed in a uniform magnetic field of 0.5 T along the beam direction, is a gaseous detector for reconstructing a charged particle's trajectory, determining its momentum and measuring its specific energy loss ($dE/dx$) for particle identification (PID). It covers full azimuth within $|\eta|<1.0$. Due to the high luminosity environment, most of reconstructed vertices using TPC tracks are from out-of-time collisions from different bunch crossings than the triggered collision. The primary vertex is chosen such that its coordinate along the beam direction ($v_{z}^{\rm{TPC}}$) is within 6 cm of the vertex $z$ position reconstructed using the VPD ($v_{z}^{\rm{VPD}}$), i.e. $\Delta v_{z}=|v_{z}^{\rm{TPC}}-v_{z}^{\rm{VPD}}|<6$ cm, as the VPD is a fast detector and thus resilient to out-of-time collisions. Such a requirement is also effective in suppressing events in which more than one collision occurs in the same bunch crossing, since in such an event the VPD picks up signals from all in-time collisions, resulting in an incorrectly reconstructed $v_{z}^{\rm{VPD}}$ that fails the $\Delta v_{z}$ requirement and is thus discarded. For the \pp\ (\pAu) data sample, the fraction of such in-time pileup events is reduced from approximately 17\% (7\%) to 2.7\% (1.3\%) after applying the $\Delta v_{z}$ cut. To further improve the vertex quality, there should be at least two tracks, which project to signals in the fast Barrel Electromagnetic Calorimeter (BEMC) \cite{STAR:2002ymp} or cross the TPC central membrane, used in reconstructing the chosen primary vertex. Additionally, $v_{z}^{\rm{TPC}}$ is required to be within $\pm100$ cm of the center of the TPC, to ensure relatively uniform TPC acceptance while maximizing the sample statistics, and the primary vertex position along the radial direction should not exceed 1.5 cm to avoid selecting collisions between the beam and the beam pipe. 

\section{Analysis details}
The invariant \jpsi\ yield per \pp\ or \pAu\ collision is obtained by dividing the efficiency and acceptance corrected \jpsi\ yield by the number of minimum-bias (MB) events in the sampled luminosity equivalent to the analyzed dimuon triggered events. The MB trigger requires a coincidence signal in the east and west VPD, a condition also included in the dimuon trigger. For the \pp\ analysis, the non-single diffractive cross section ($\sigma_{pp}^{\rm{NSD}}=30.0\pm2.4$ mb \cite{STAR:2012nbd}) at 200 GeV is used to convert the measured invariant yield to a cross section. The equivalent number of MB events is calculated using the recorded number of MB events corrected for the efficiency and prescale factor ($f_{\rm{prescale}}$) of the MB trigger, where 1/$f_{\rm{prescale}}$ is the fraction of MB triggered events randomly selected to be written on tape given the limited STAR data acquisition bandwidth.
 
\subsection{\jpsi\ signal extraction}
TPC tracks are chosen only if their Distance of Closest Approach (DCA) to the primary vertex is less than 3 cm. The primary vertex is then included in a refit of the track to improve the momentum resolution. To ensure good momentum and $dE/dx$ resolutions, the number of TPC space points used for track reconstruction and $dE/dx$ calculation are at least 20 and 15, respectively. The ratio of the number of TPC space points used for track reconstruction to the maximum possible number of space points along the track trajectory should be no less than 0.52 to reject split tracks. 

\begin{figure}[htbp]
\begin{minipage}{1.0\linewidth}
\centerline{\includegraphics[width=0.7\linewidth]{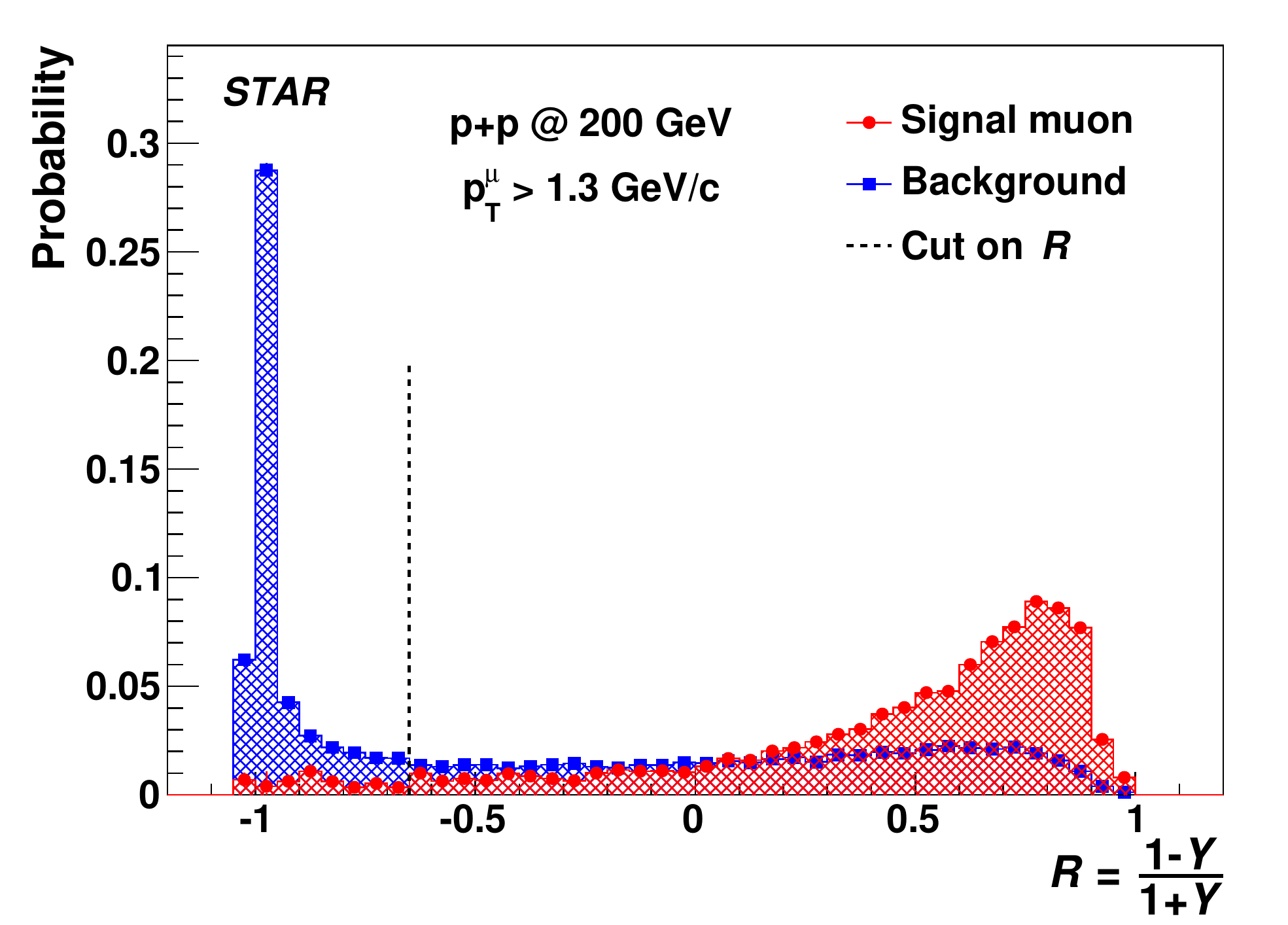}}
\end{minipage}
\caption[]{$R$ distribution for signal muons (circles) and background (squares) in \pp\ analysis. The dashed vertical line indicates the cut to select muon candidates.}
\label{fig:LRpp}
\end{figure}

Tracks are further extrapolated radially to the middle of MTD modules, located at varying distances between 392.8 cm and 418.9 cm from the center of STAR, and matched to the closest MTD hits. If more than one track is matched to the same hit, the closest track is chosen.  Once a track-hit association is established, the track is identified as a muon candidate when the following two criteria are satisfied: i) the associated MTD hit contributes to the dimuon trigger; ii) the pair survives the cut on a Likelihood Ratio ($R$) \cite{Huang:2016dbm}. The $R$ variable is defined as the following:
\begin{equation}
R=\frac{1-Y}{1+Y},\  Y = \displaystyle \prod_{i} \frac{\mathrm{pdf}_{i}^{\mathrm{bkg}}}{\mathrm{pdf}_{i}^{\mathrm{sig}}}
\label{eq:LR}
\end{equation}
where $i$ stands for the five discriminating variables used, i.e., DCA, \nsigmapi, \deltaznorm, \mbox{\deltayqnorm}, \deltatof, and $\mathrm{pdf}_{i}^{\mathrm{sig}}$, and $\mathrm{pdf}_{i}^{\mathrm{bkg}}$ are the probability distribution functions of each variable for signal muons and background particles. The normalized energy loss is defined as 
$\nsigmapi=\frac{\ln(dE/dx)_{\rm{measured}}-\ln(dE/dx)_{\rm{theory}}^{\pi}}{\sigma(\ln(dE/dx))}$. Here $(dE/dx)_{\rm{measured}}$ is the measured energy loss in the TPC, $(dE/dx)_{\rm{theory}}^{\pi}$ is the expected energy loss for a pion based on the Bichsel formalism \cite{Bichsel:2006cs}, and $\sigma(\ln(dE/dx))$ is the resolution of the $\ln(dE/dx)$ measurement. The shape of the \nsigmapi\ distribution is expected to be independent of track \pT. \deltaz\ and \deltay\ are the position differences between the projected track trajectories on the MTD and the associated MTD hits along the $z$ and azimuthal directions, and \deltay\ is multiplied by the track's electric charge ($q$) to eliminate the difference between positively and negatively charged particles. These two variables are divided by their resolutions to remove their \pT\ dependences. Finally, \deltatof\ represents the difference between the measured and expected flight time for a muon flying from the primary vertex to the MTD. The measured flight time is based on the MTD and VPD information, while the expected value is obtained by dividing the track flight length by its speed. The probability distribution functions for signal muons and background particles are extracted from data. The signal muon sample is obtained from muon candidate pairs of unlike charge signs (``unlike-sign") around the \jpsi\ mass range ($3.0<M_{\mu^{+}\mu^{-}}<3.2$ \gevtwo) after subtracting the muon candidate pairs of like charge signs (``like-sign"), where the muon candidates are identified without applying the PID cut in question. The like-sign pairs of muon candidates are used as background particles. The value of $R$ is expected to approach 1 for signal muons, and -1 for background. For example, the distributions of $R$ for signal muons and background particles above 1.3 \gev\ are shown in Fig. \ref{fig:LRpp} from the \pp\ data sample. A clear separation between signal and background is seen. 
The optimal cut of $R>-0.65$, independent of muon \pT, is determined by maximizing the \jpsi\ signal significance in the entire \pT\ range, and is shown as the vertical dashed line in Fig. \ref{fig:LRpp}. The optimal cut is determined to be $R>-0.01$ for the \pAu\ analysis.

\begin{figure}[htbp]
\begin{minipage}{0.49\linewidth}
\centerline{\includegraphics[width=0.95\linewidth]{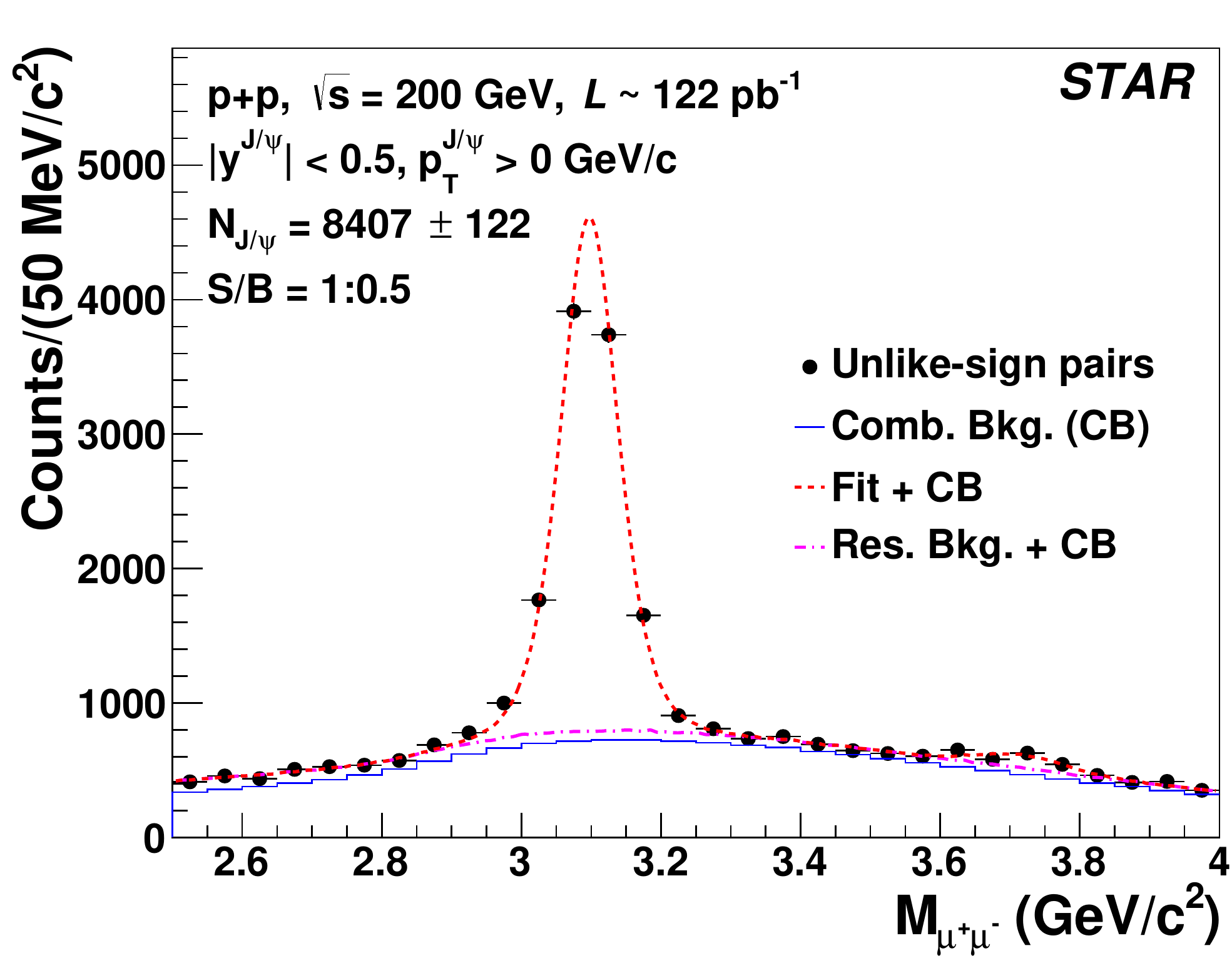}}
\end{minipage}
\begin{minipage}{0.49\linewidth}
\centerline{\includegraphics[width=0.95\linewidth]{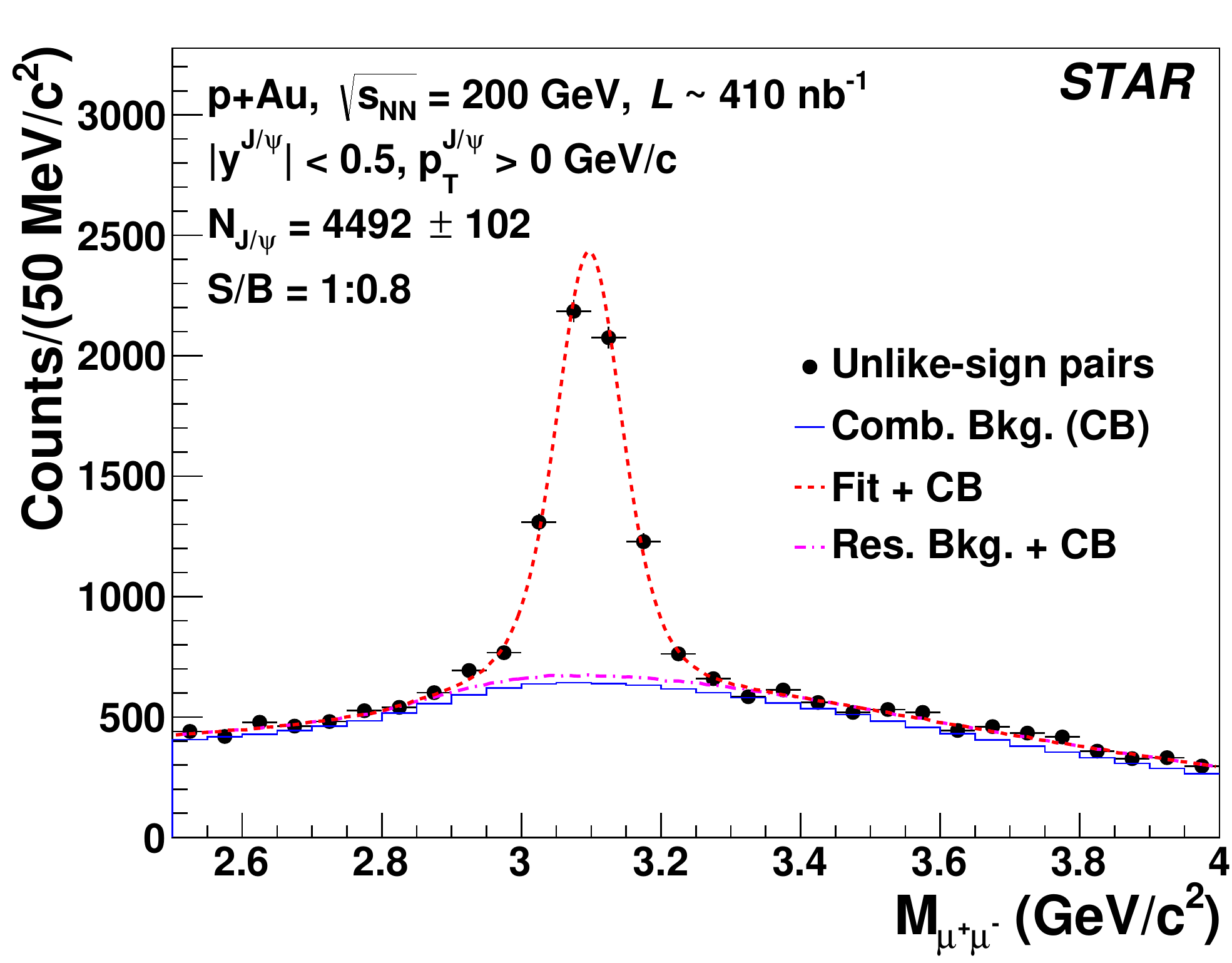}}
\end{minipage}
\caption[]{Invariant mass distributions of unlike-sign muon pairs (filled circles) in \pp\ (left) and \pAu\ (right) collisions. These distributions, after subtracting the combinatorial background (CB, blue histograms), are fit with a Student's $t$ function describing the signal and a first-order polynomial function representing the residual background. The resulting combined (dashed lines) and residual background (dot-dashed lines) fits are also shown with the combinatorial background. Horizontal and vertical bars around data points depict bin width and statistical errors, respectively. }
\label{fig:InvMass}
\end{figure}

The invariant mass distributions for unlike-sign muon candidate pairs, integrated over \pT, are shown in Fig.~\ref{fig:InvMass} as filled circles for \pp\ (left) and \pAu\ (right) events. For each unlike-sign pair, one of the muon candidates should have a \pT\ above 1.5 \gev, while the other above 1.3 \gev. The pair rapidity is within $|y|<0.5$. 
The combinatorial background is estimated using like-sign TPC track pairs without matching to the MTD, which are scaled to the invariant mass distributions of like-sign muon candidate pairs and corrected for the MTD acceptance difference for like-sign and unlike-sign pairs. Such an acceptance difference is evaluated using the ratio of unlike-sign to like-sign muon candidate pairs from mixed events, i.e. the two muons in a pair are taken from different events, using 200 GeV Au+Au collisions recorded in 2014 \cite{Adam:2019rbk}. To extract the raw \jpsi\ yield, the unlike-sign invariant mass distribution, with the combinatorial background subtracted, is fit with the sum of a Student's $t$ function representing the \jpsi\ signal and a first-order polynomial function describing the residual background, using the minimum $\chi^{2}$ method. The normality parameter in the Student's $t$ function, which determines to what extent its tail is enhanced with respect to a Gaussian function, is fixed according to simulations. Fit results for signal plus residual background and residual background only are shown as dashed and dot-dashed lines in Fig.~\ref{fig:InvMass} with the combinatorial background added back. This procedure is applied to the \jpsi\ signal extraction for \pT\ below 4 \gev. For \pT\ greater than 4 \gev, where the signal-to-background ratio increases and the statistics decreases, the unlike-sign distributions, without background subtraction, are directly fit with a Student's $t$ function plus a first-order polynomial function using the maximum likelihood method. The signal-to-background ratios listed in Fig.~\ref{fig:InvMass} are calculated within \mbox{$3.0<M_{\mu^{+}\mu^{-}}<3.2$ \gevtwo}.

\subsection{Efficiency and acceptance correction}
The TPC tracking efficiency and acceptance are evaluated through an embedding procedure. Simulated $\jpsi\rightarrow\mu^{+}\mu^{-}$ processes are propagated through the STAR detector simulation using the GEANT3 package \cite{Brun:1987ma}. They are then mixed with randomly sampled dimuon triggered real events, and reconstructed the same way as real data. The embedded \jpsi\ is assumed to have zero polarization \cite{Adam:2020dcj}. The TPC tracking efficiency is 86\% (85.5\%) independent of muon \pT\ above 1.3 \gev\ for \pp\ (\pAu) data. An additional inefficiency observed in the data for one of the TPC sectors is applied to the embedding sample.

The MTD matching efficiency for muons includes contributions from the MTD acceptance and the response efficiency of each MTD module, which is defined as the probability for a track to generate a hit in the module when extrapolated to the active volume. The MTD acceptance is simulated in the aforementioned embedding sample, while the response efficiencies are assessed using cosmic ray data. For MTD modules residing at the bottom hemisphere of the STAR detector, the response efficiencies as a function of muon \pT\ are obtained by extrapolating cosmic ray tracks to the active volume and finding the fraction of tracks matched to MTD hits. The average response efficiency of all bottom modules is also used as a template for MTD modules located at the top hemisphere, for which the cosmic rays travel from outside in, which is the opposite from the real collision data. The absolute scale of the template is determined by matching to the response efficiency of each top module at $\pT> 5$ \gev, where the efficiency reaches a plateau. The extracted response efficiency for each MTD module is then applied to the embedding sample. The resulting MTD matching efficiencies for bottom modules evaluated using the embedding sample and the cosmic ray data are consistent with each other. To take into account the residual differences, the ratio between the average of the two and the matching efficiency from embedding is used as an additional scale factor for obtaining the final MTD matching efficiency for all the modules. This procedure applies equally to both \pp\ and \pAu\ analyses.

The MTD trigger efficiency consists of three components: trigger electronics efficiency, trigger timing window cut efficiency and trigger patch configuration efficiency. The first two components are evaluated using MB triggered event samples for which only the coincidence signal in east and west VPD is required. The probability for a muon candidate to generate a correct signal in the trigger electronics and pass the trigger timing window cut is found to be close to 100\% for both \pp\ and \pAu\ data. The third component arises from the fact that the dimuon trigger requires signals from distinct trigger patches \cite{Adam:2019rbk} while the muon daughters from high-\pT\ \jpsi\ decays are highly boosted and could hit the same MTD trigger patch. Since this component is driven by the MTD geometry and the \jpsi\ decay kinematics, the embedding sample is utilized. The resulting efficiency is mostly 100\% until \jpsi\ \pT\ of 5 \gev, and decreases to about 95\% at 8-10 \gev.

The muon PID efficiency associated with the cut on $R$ is estimated using a tag-and-probe method based on real data. For each unlike-sign pair of TPC tracks matched to the MTD, one muon is randomly selected as the {\it tag} muon while the other the {\it probe} muon. For the tag muon, a strict cut of $R>0.25$ is applied to increase the signal-to-background ratio. For the probe muon, two cases are tried, i.e., no cut on $R$ and the default cut on $R$. The \jpsi\ counts in each probe muon \pT\ bin are extracted for the two cases, and the ratio is parametrized as the muon PID efficiency. In the \pp\ analysis, the muon PID efficiency increases from 90\% at 1.3 \gev\ to 98\% above 5 \gev, while it increases from 69\% at 1.3 \gev\ to 96\% above 5 \gev\ for \pAu\ analysis due to the tighter cut applied. 

The VPD trigger and vertex finding efficiencies are obtained by embedding PYTHIA \cite{Sjostrand:2006za,Sjostrand:2007gs} (HIJING \cite{Wang:1991hta}) events, after passing through the GEANT simulation of the STAR detector, into zero-bias \pp\ (\pAu) events. The zero-bias events were taken without any trigger requirement at random times. For the \pp\ analysis, both the MB PYTHIA events and PYTHIA events containing a \jpsi\ within $|y| < 0.5$ are used for embedding. The former is needed for calculating the equivalent number of MB events corresponding to the analyzed dimuon triggered events. Two different PYTHIA configurations: i) PYTHIA 6.4.28 \cite{Sjostrand:2006za} plus the Perugia2012 tune \cite{Skands:2010ak}; ii) PYTHIA 8.1.62 \cite{Sjostrand:2007gs} with the STAR heavy flavor tune as detailed in the appendix, are used as they bracket the measured multiplicity distribution of \jpsi\ events \cite{Adam:2018jmp}. To account for the apparent differences between data and PYTHIA, event multiplicity distributions for both MB \cite{Ansorge:1988kn} and \jpsi\ events \cite{Adam:2018jmp} are then used to weight the embedding samples. The VPD efficiency for events containing a \jpsi\ decreases with increasing \jpsi\ \pT, due to the decreased amount of energy available for producing particles in the VPD acceptance in these events. The average efficiencies of the two PYTHIA configurations are taken as the central values, while half the difference of the two is taken as a source of systematic uncertainties. For the \pAu\ analysis, a similar procedure is used except that the HIJING event generator \cite{Wang:1991hta} is employed. Since no quarkonium production is implemented, HIJING events containing a $D^{0}$ meson within $|y|<0.5$ are used for embedding, which is validated by the good agreement seen between the efficiencies extracted from embedding \jpsi\ and $D^{0}$ PYTHIA events. The event multiplicity as a function of pseudorapidity in HIJING \pAu\ events is compared to the PHOBOS measurement for \dAu\ collisions \cite{Alver:2010ck} scaled by the difference in \npart\ between \pAu\ and \dAu\ collisions. Here \npart\ refers to the number of participating nucleons in a \pAu\ or \dAu\ collision. These results agree for both mid-rapidity and the $p$-going side. However, on the Au-going side HIJING significantly underpredicts the particle multiplicity, and therefore the VPD efficiency in this side is assumed to be 100\% as an upper limit. The average VPD efficiency in the Au-going side ($\sim91$\%) from the default HIJING and the upper limit is taken as the central value.

\subsection{Systematic uncertainties}
Systematic uncertainties due to different aspects of the analysis procedure are evaluated. For the signal extraction, the following variations are evaluated. When determining the combinatorial background scaling, a first-order, instead of a second-order, polynomial function is used and the fitting range is varied from [2.5, 4.0] \gevtwo\ to [2.3, 4.2] \gevtwo\ and [2.55, 3.9] \gevtwo. When fitting the invariant mass distributions, the fitting range is changed by 0.3 to 0.6 \gevtwo\ depending on the \jpsi\ \pT, a second-order polynomial function is used for fitting the residual background, and the binning for the invariant mass distribution is varied from 50 (80) MeV/$c^{2}$ to 20 (50) MeV/$c^{2}$ for $\pT<(>)\ 3.5$ \gev. The simulation uncertainty in the normality parameter is also taken into account. Furthermore, \jpsi\ yields are extracted by counting the unlike-sign muon candidate pairs around \jpsi\ mass after background subtraction. The maximum deviations from the default cases are taken as the uncertainties. The uncertainty in the TPC tracking efficiency is evaluated by changing the number of TPC space points used for track reconstruction and $dE/dx$ calculation from 20 and 15 to 25 and 20 or 15 and 10 simultaneously in data analysis and efficiency estimation, and the changes in the final results are used as the systematic uncertainty. The resulting tracking efficiency uncertainty is 4\% independent of \jpsi\ \pT\ for both \pp\ and \pAu\ analyses. In terms of the MTD matching efficiency, its uncertainty includes three contributions: i) statistical precision of the cosmic ray data used to determine the MTD response efficiencies; ii) the uncertainty arising from using the response efficiency template for the top MTD modules which is estimated as the average absolute differences between the response efficiency template and the actual response efficiencies for bottom MTD modules; iii) half of the difference in the matching efficiencies between using cosmic ray data and the embedding sample. The MTD matching efficiency uncertainty decreases from 6.0\% at a \jpsi\ \pT\ of 1 \gev\ to 1.9\% at 10 \gev. For both the muon PID and MTD trigger efficiencies, their uncertainties are driven by the statistical precision of the data-driven methods used. Systematic uncertainties associated with the VPD trigger and vertex finding efficiencies include the following three contributions: statistical precision of the embedding sample, deviations between different PYTHIA or HIJING configurations and the central value, and the variation in the VPD response efficiency from 100\% to 90\%. The former is uncorrelated among different \pT\ bins, while the latter two are independent of \pT\ and included in the global uncertainties. The impact of remaining pileup contribution to the \jpsi\ yield is estimated to be 2.7\% and 1.3\% for \pp\ and \pAu\ analyses, respectively, and assigned as a source of uncertainty. Finally, an 8\% uncertainty on $\sigma_{pp}^{\rm{NSD}}$ is added \cite{STAR:2012nbd}. For the \rpa\ measurement, the uncertainties of the TPC tracking efficiency, MTD matching efficiency and pileup contribution mostly cancel, while other sources of uncertainties are uncorrelated between \pp\ and \pAu\ analyses and thus added in quadrature. All the individual sources of uncertainties are listed in Table \ref{tb:SysAll}, along with the total uncertainties obtained by adding individual ones in quadrature. Global uncertainties, referred to in later sections, include those from the \pT-independent part in VPD trigger and vertex finding efficiencies, pileup contribution, $\sigma_{pp}^{\rm{NSD}}$ and \ncoll.

\begin{table}[htbp]
\centering
\begin{tabular}{|c|c|c|c|} 
 \hline
Uncertainty source & \pp\  & \pAu\ & \rpa\\ 
\hline
Signal extraction &  1.1 - 8.5\%& 2.1 - 6.0\% & 2.7-10.4\% \\
\hline
TPC tracking & 4\% & 4\% &  cancelled \\
\hline
MTD matching & 1.9 - 5.5\% & 1.9 - 6.0\%& negligible \\
\hline
Muon PID & 0.9 - 1.2\% & 3.0 - 4.8\% & 3.1 - 4.9 \%\\
\hline
MTD trigger & 1.4\% & 1.4\% & 2\% \\
\hline
VPD trigger and vertex finding  & 9.3 - 15.2\% & 1.7 - 11.0\% & 9.4 - 18.8\% \\
\hline
Pileup & 2.7\% & 1.3\% & 1.4\%\\
\hline
$\sigma_{pp}^{\rm{NSD}}$ & 8\% & - & 8\%\\
\hline
\ncoll\ & - & - & 6.4\%\\
\hline
\hline
Total & 13.5 - 20.0\% & 6.9 - 13.9\% & 14.9\%-24.2\%\\
\hline
\end{tabular}
\caption{List of individual and total systematic uncertainties. A range is given if the uncertainty varies with \jpsi\ \pT. }
\label{tb:SysAll}
\end{table}

\section{Results and discussions}
\begin{figure}[htbp]
\begin{minipage}{1.0\linewidth}
\centerline{\includegraphics[width=0.6\linewidth]{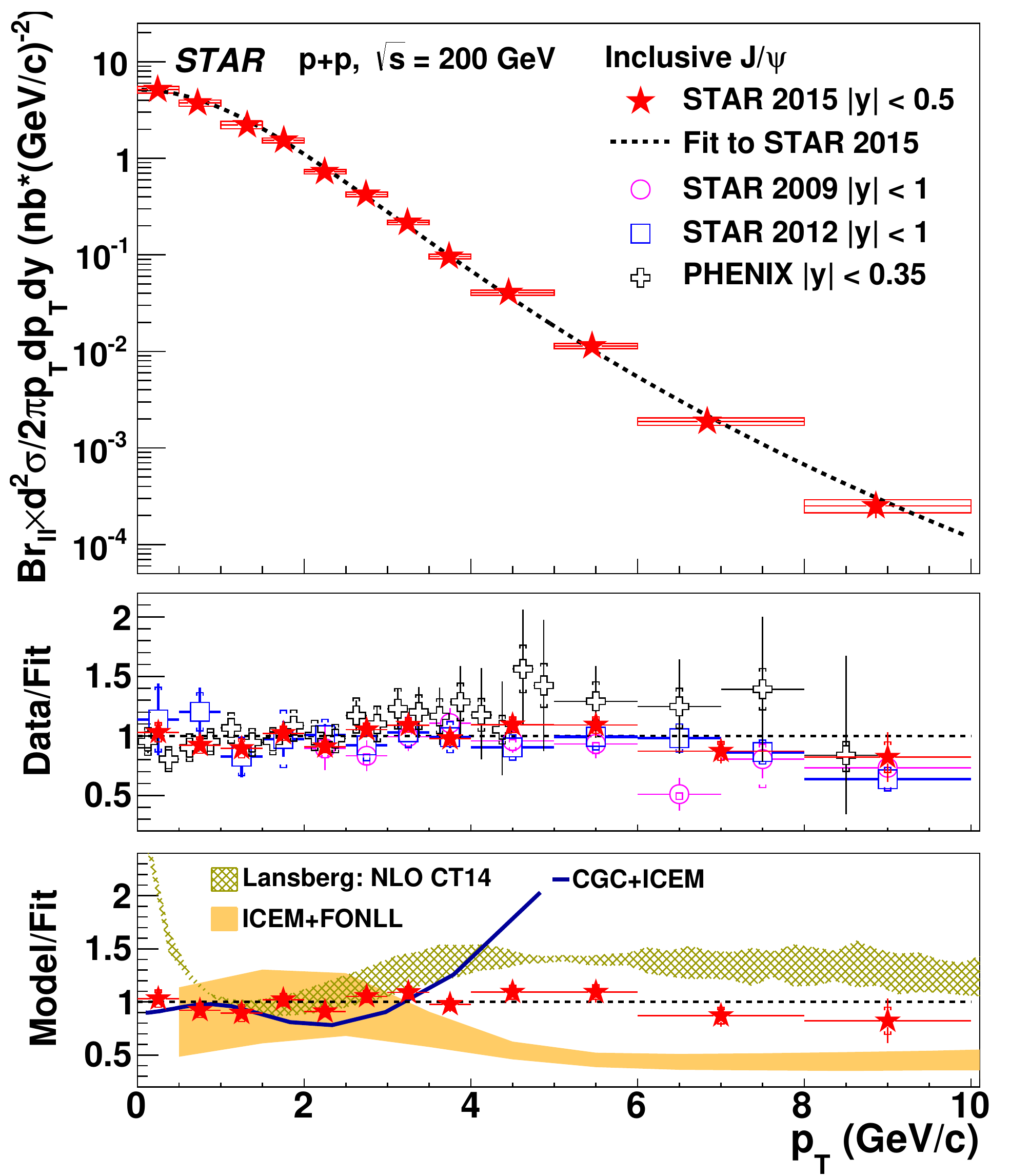}}
\end{minipage}
\caption[]{Top: inclusive \jpsi\ cross section (red stars) as a function of \pT\ in \pp\ collisions at \sqrts\ = 200 GeV, with a fit to the result shown as the dashed line. The vertical error bars, smaller than the marker size, and open boxes around data points represent statistical errors and systematic uncertainties. The horizontal error bars indicate the bin width. Middle: ratios of current and previous measurements \cite{Adamczyk:2012ey, Adare:2011vq, Adam:2018jmp} to the fit, with correction factors applied to account for different rapidity coverages. The vertical error bars and brackets around data points represent statistical errors and systematic uncertainties. Global uncertainties are not shown for all the measurements. They are 12.5\% for this analysis, 8.1\% for STAR 2009 result \cite{Adamczyk:2012ey}, 10\% (8.5\%) for STAR 2012 result at $\pT < 1.5$ \gev\ ($\pT > 1.5$ \gev) \cite{Adam:2018jmp} and 10\% for the PHENIX measurement \cite{Adare:2011vq}. Bottom: ratios of current measurement and different model calculations \cite{Ma:2016exq, Ma:2017rsu, Lansberg:2016deg, Kusina:2017gkz, Shao:2015vga, Shao:2012iz} to the fit. Systematic uncertainties on data points are smaller than the marker size.}
\label{fig:JpsiInpp}
\end{figure}

The differential cross section of inclusive \jpsi\ times the branching ratio within $|y|<0.5$ in \pp\ collisions at \sqrts\ = 200 GeV is shown in the top panel of Fig.~\ref{fig:JpsiInpp} as a function of \pT. The data points are placed at the \pT\ positions whose yields are equal to the average yields of the corresponding bins \cite{Lafferty:1994cj}. For this purpose, the following empirical function is used to fit the differential cross section as a function of \pT\ iteratively: 
\begin{equation}
f(\pT) = A\times\pT\times(1+(\pT/B)^{2})^{C},
\label{eq:jpsifit}
\end{equation}
where $A$, $B$ and $C$ are free parameters. The integrated \jpsi\ cross section per unity rapidity is: 
\begin{equation}
\mathrm{Br}_{\mu\mu}\frac{\mathrm{d}\sigma_{\jpsi}}{\mathrm{d}y}|_{y=0} = 43.9\pm0.7 (stat.) \pm 6.1 (syst.)\ \mathrm{nb}
\label{eq:jpsixsec}
\end{equation}

To facilitate the comparison of this measurement to previous publications, Eq. \ref{eq:jpsifit} is used to fit the differential \jpsi\ cross section, and the fit result is shown as the dashed line in the top panel of Fig.~\ref{fig:JpsiInpp}. Ratios of the current and previous results \cite{Adamczyk:2012ey, Adare:2011vq, Adam:2018jmp} to the fit function are shown in the middle panel of Fig.~\ref{fig:JpsiInpp}, with correction factors applied to convert the measurements of different rapidity coverages into a common range of $|y|<0.5$. The correction factors are based on calculations of the Improved Color Evaporation Model (ICEM) \cite{Ma:2016exq}, as described below, showing that the invariant \jpsi\ yields within $|y|<0.5$  and $|y| < 0.35$ agree with each other within 2\% while the yield within $|y|<0.5$ is about 6\% (11\%) larger at 0 (10) \gev\ than that within $|y| < 1$. The different STAR analyses are consistent with one another, while the new measurement improves the precision below 1 \gev. On the other hand, while the PHENIX measurement \cite{Adare:2011vq} is also consistent with this analysis within uncertainties, its central values are systematically higher above 2.5 \gev. Ratios of three model calculations to the fit are shown in the lower panel of Fig.~\ref{fig:JpsiInpp}, and compared to data. The ICEM determines the transition probability from $c\bar{c}$ pairs to \jpsi\ by fitting to \jpsi\ measurements from previous publications \cite{Ma:2016exq}. The associated uncertainties arise from varying the charm quark mass ($m_{c}$), factorization and renormalization scales. For the calculation labeled as ``CGC+ICEM", it utilizes the CGC framework to obtain the $c\bar{c}$ production cross section and the ICEM for hadronization. Since the variation in $m_{c}$ between 1.3 and 1.4 \gevtwo\ leads to negligible differences in this work, only the results with $m_{c}=1.3$ \gevtwo\ are compared to data. The model calculation from Lansberg is based on the CT14 proton PDF at Next-to-Leading Order (NLO) \cite{Dulat:2015mca} and its uncertainties are dominated by variations in the factorization scale. Both ``CGC+ICEM" and Lansberg calculations are tuned to previous inclusive \jpsi\ cross section measurement at 200 GeV \cite{Adare:2011vq}, and therefore can be directly compared to this measurement. The ICEM calculation includes only prompt \jpsi, i.e. directly produced \jpsi\ plus those from decays of excited charmonium states. The contribution to \jpsi\ from $b$-hadrons is calculated at Fixed Order plus Next-to-Leading Logarithms (FONLL) \cite{Cacciari:1998it, Cacciari:2001td} and added to the ICEM result. The $b$-hadron feeddown contribution increases from less than 1\% below 1 \gev\ to approximately 9\% at 10 \gev. Uncertainties of the FONLL calculation are added in quadrature to ICEM model uncertainties. The ICEM and ``CGC+ICEM" results agree with the data within uncertainties up to about 3.5 \gev\ before diverging from the data. The Lansberg calculation is consistently above the data, especially for $\pT < 0.5$ \gev. 

\begin{figure}[htbp]
\begin{minipage}{1.0\linewidth}
\centerline{\includegraphics[width=0.6\linewidth]{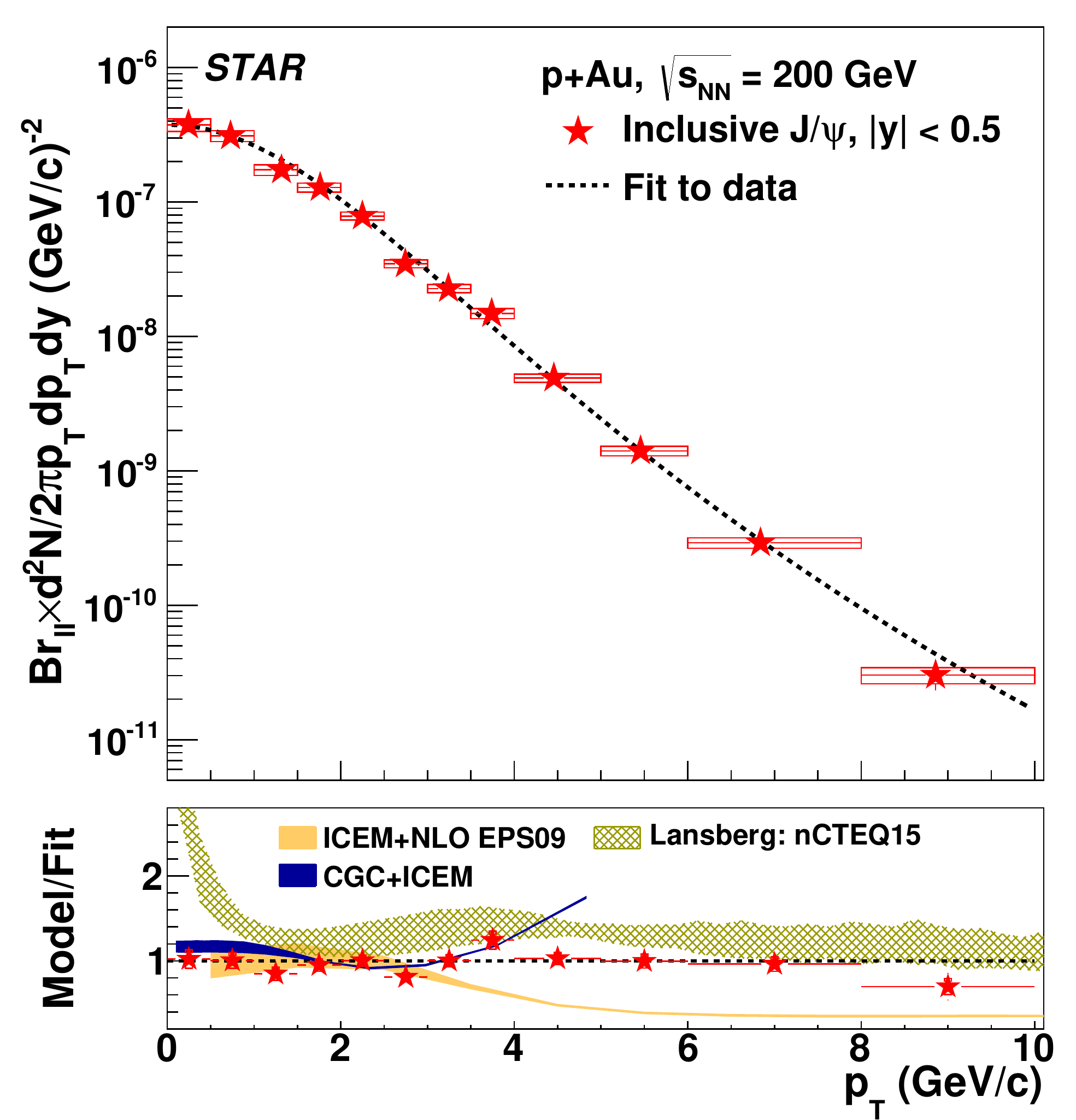}}
\end{minipage}
\caption[]{Top: inclusive \jpsi\ yield (red stars) as a function of \pT\ in \pAu\ collisions at \sqrtsNN\ = 200 GeV, with a fit to the result shown as the dashed line. The vertical error bars, smaller than the marker size, and open boxes around data points represent statistical errors and systematic uncertainties. The global uncertainty of 1.5\% is not shown. The horizontal error bars indicate the bin width. Bottom: ratios of data and different model calculations \cite{Ma:2016exq, Ma:2017rsu, Lansberg:2016deg, Kusina:2017gkz, Shao:2015vga, Shao:2012iz} to the fit. Systematic uncertainties on data points are smaller than the marker size.}
\label{fig:JpsiInpa}
\end{figure}

The inclusive \jpsi\ yield times the branching ratio as a function of \pT\ in \pAu\ collisions at \sqrtsNN\ = 200 is shown in Fig.~\ref{fig:JpsiInpa}, top panel. Similarly, the yield is fit with Eq. \ref{eq:jpsifit}, which is shown as the dashed line. Ratios of the data and different model calculations to the fit are shown in the bottom panel of Fig.~\ref{fig:JpsiInpa}.
The ICEM utilizes the NLO EPS09 nPDF \cite{Eskola:2009uj}, while the CGC+ICEM approach directly calculates the $c\bar{c}$ production cross section in \pAu\ collisions based on the CGC formalism. The uncertainties for the former arise from the nPDF uncertainties, while for the latter the main contributions are the variations of the average momentum of soft color exchanges and the scale factor between the saturation scales for proton and Au nucleus. In the Lansberg calculation, the nCTEQ15 nPDF at NLO \cite{Kovarik:2015cma}, constrained by the \jpsi\ measurements at the LHC, is used \cite{Lansberg:2016deg, Kusina:2017gkz, Shao:2015vga, Shao:2012iz}. The systematic uncertainty band of the Lansberg calculation includes the nPDF uncertainty at 68\% confidence level as well as variations on the factorization scale. Very similar results, not shown here, are obtained using the EPPS16 nPDF at NLO \cite{Eskola:2016oht} within the same framework. The comparison between data and model calculations is similar to that seen for \pp\ collisions, even though the contribution of b-hadron decayed \jpsi\ is not included in the model calculations for \pAu\ collisions. 

\begin{figure}[htbp]
\begin{minipage}{1.0\linewidth}
\centerline{\includegraphics[width=0.6\linewidth]{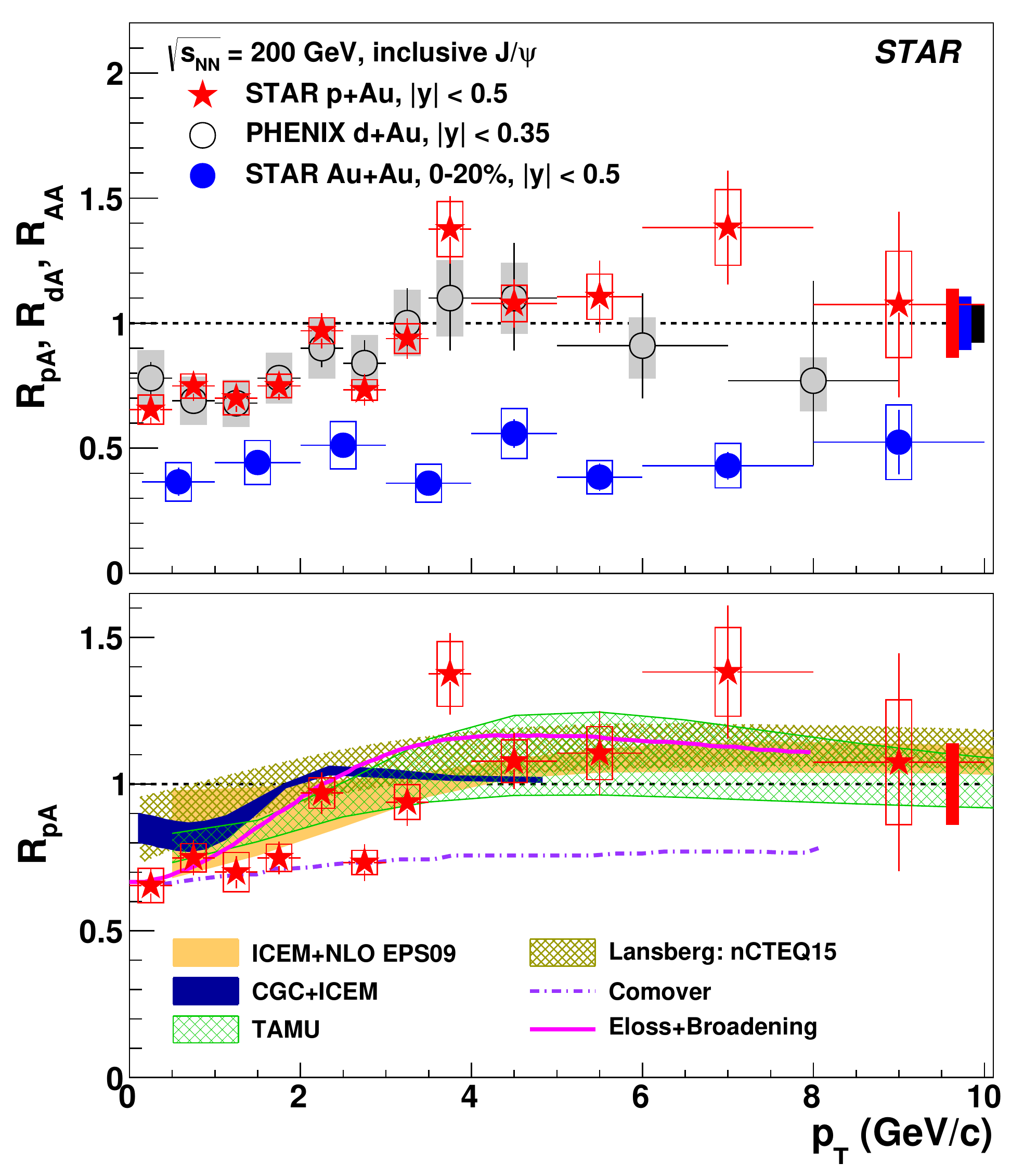}}
\end{minipage}
\caption[]{Top: inclusive \jpsi\ \rpa\ (filled stars) and \rda\ (open circles) \cite{Adare:2012qf} as a function of \pT\ compared to the \jpsi\ \RAA\ (filled circles) measured in 0-20\% central \AuAu\ collisions at \sqrtsNN\ = 200 GeV. The error bars and open boxes around data points represent statistical errors and systematic uncertainties, while the filled boxes at unity display the global uncertainties for each data set. Bottom: comparison of \jpsi\ \rpa\ to different model calculations \cite{Arleo:2013zua, Ferreiro:2014bia, Du:2015wha, Du:2018wsj, Ma:2016exq, Ma:2017rsu, Lansberg:2016deg, Kusina:2017gkz, Shao:2015vga, Shao:2012iz}.}
\label{fig:Jpsirpa}
\end{figure}

Following Eq. \ref{eq:rpa}, the inclusive \jpsi\ \rpa, quantifying the CNM effects on \jpsi\ production in 200 GeV \pAu\ collisions, is calculated and shown as a function of \pT\ in the top panel of Fig.~\ref{fig:Jpsirpa}. The global uncertainty is shown as the filled box at unity.
A suppression of approximately 30\% is seen below 2 \gev, which gradually goes away as \pT\ increases. For \pT\ above 3 \gev, the \jpsi\ \rpa\ becomes consistent with unity, indicating little CNM effects on the \jpsi\ production in this kinematic range. The inclusive \jpsi\ \rda\ \cite{Adare:2012qf} in \dAu\ collisions at \sqrtsNN\ = 200 GeV is shown as open circles for comparison. It agrees with the \jpsi\ \rpa\ within uncertainties, indicating that the CNM effects in \pAu\ and \dAu\ collisions are similar. It is worth noting that the new \jpsi\ \rpa\ has a better precision than the published \jpsi\ \rda\ in the entire \pT\ range. Also shown in the panel as filled circles are the \jpsi\ \RAA\ measured in 0-20\% central \AuAu\ collisions at \sqrtsNN\ = 200 GeV \cite{Adam:2019rbk}, in which the \jpsi\ yield is seen to be suppressed over the entire \pT\ range. For $\pT<2$ \gev, the CNM effects contribute significantly to the \jpsi\ suppression seen in heavy-ion collisions, while for $\pT>3$ \gev, the dissociation effect arising from the presence of the QGP medium is mainly responsible for the strong suppression. Different model calculations are shown in the bottom panel of Fig.~\ref{fig:Jpsirpa}, and compared to the data. The ICEM and Lansberg calculations include only nPDF effects based on EPS09 \cite{Eskola:2009uj} and nCTEQ15 \cite{Kovarik:2015cma} parameterizations, respectively. The TAMU model extends the transport model for heavy-ion collisions to \pAu\ collisions \cite{Du:2015wha, Du:2018wsj}. In this model, the NLO EPS09 nPDF is utilized \cite{Eskola:2009uj}, and the short-lived hot medium modifies the observed \jpsi\ yields in \pAu\ collisions through both dissociation and recombination. Uncertainties of this calculation includes nPDF uncertainties, variation of the broadening parameter for incorporating the Cronin effect, and uncertainties in the formation times for both the QGP and \jpsi\ meson. In another model, shown as the solid line and labeled as ``Eloss+Broadening", interactions between fast-moving color-octet $c\bar{c}$ pairs in the nucleus rest frame and the cold nuclear medium induce both radiative energy loss and \pT-broadening \cite{Arleo:2013zua}. The latter is responsible for the \jpsi\ enhancement above 2.5 \gev. A comover model, introducing breakup of \jpsi\ mesons through interactions with final state particles traveling along with \jpsi, is shown as the dot-dashed line in the panel \cite{Ferreiro:2014bia}. The nPDF effect is also included in the comover model, based on leading-order EPS09 parameterization. All the model calculations are consistent with data within theoretical and experimental uncertainties. It is worth noting that the comover model underpredicts data above 3.5 \gev\ by 2.3$\sigma$.

\section{Summary}
In summary, the inclusive \jpsi\ yields within $|y|<0.5$ are measured in \pp\ and \pAu\ collisions at \sqrtsNN\ = 200 GeV through the dimuon decay channel using the STAR experiment. Both data samples were taken with the MTD dimuon trigger in 2015. The differential \jpsi\ cross section in \pp\ collisions is consistent with previous results, and the data precision for $\pT<1$ \gev\ is significantly improved for STAR measurements. For the first time, the \jpsi\ \rpa\ at mid-rapidity ($|y|<0.5$) is measured within $0<\pT<10$ \gev\ to quantify the CNM effects experienced by the \jpsi\ meson in \pAu\ collisions. It increases from 0.65 at 0-0.5 \gev\ to be consistent with unity above 3 \gev. Comparison to a similar measurement in 0-20\% central \AuAu\ collisions at \sqrtsNN\ = 200 GeV confirms that the observed large suppression of the \jpsi\ yield above 3 \gev\ is mostly caused by hot medium effects, providing strong evidence of the QGP formation in these collisions. Model calculations including different underlying physics mechanisms can qualitatively describe the data within uncertainties. The new measurements presented in this paper provide an improved reference for interpreting similar measurements in 200 GeV \AuAu\ collisions, and will further constrain model calculations of the CNM effects for \jpsi\ at RHIC. 

\section*{Acknowledgements}
We thank the RHIC Operations Group and RCF at BNL, the NERSC Center at LBNL, and the Open Science Grid consortium for providing resources and support.  This work was supported in part by the Office of Nuclear Physics within the U.S. DOE Office of Science, the U.S. National Science Foundation, the Ministry of Education and Science of the Russian Federation, National Natural Science Foundation of China, Chinese Academy of Science, the Ministry of Science and Technology of China and the Chinese Ministry of Education, the Higher Education Sprout Project by Ministry of Education at NCKU, the National Research Foundation of Korea, Czech Science Foundation and Ministry of Education, Youth and Sports of the Czech Republic, Hungarian National Research, Development and Innovation Office, New National Excellency Programme of the Hungarian Ministry of Human Capacities, Department of Atomic Energy and Department of Science and Technology of the Government of India, the National Science Centre of Poland, the Ministry  of Science, Education and Sports of the Republic of Croatia, RosAtom of Russia and German Bundesministerium f\"ur Bildung, Wissenschaft, Forschung and Technologie (BMBF), Helmholtz Association, Ministry of Education, Culture, Sports, Science, and Technology (MEXT) and Japan Society for the Promotion of Science (JSPS).

\section*{Appendix}
\label{app:pythia}
The settings of the STAR heavy flavor tune, based on the default PYTHIA  8.1.62 \cite{Sjostrand:2007gs} settings and LHAPDF 6.1.4 \cite{Buckley:2014ana}, are listed below:
\begin{itemize}
\item PYTHIA8::Set(``PDF:useLHAPDF = on");
\item PYTHIA8::Set(``PDF:LHAPDFset = MRSTMCal.LHgrid");
\item PYTHIA8::Set(``PDF:extrapolateLHAPDF = on");
\item PYTHIA8::Set(``SigmaProcess:renormScale2 = 3");
\item PYTHIA8::Set(``SigmaProcess:factorScale2 = 3");
\item PYTHIA8::Set(``SigmaProcess:renormMultFac = 2"); 
\item PYTHIA8::Set(``SigmaProcess:factorMultFac = 2");
\item PYTHIA8::Set(``PartonLevel:MI = on");
\item PYTHIA8::Set(``PartonLevel:ISR = on");
\item PYTHIA8::Set(``BeamRemnants:primordialKT = on");
\item PYTHIA8::Set(``PartonLevel:FSR = on");
\item PYTHIA8::Set(``StringFlav:mesonCvector = 1.5");
\item PYTHIA8::Set(``StringFlav:mesonBvector = 3");
\item PYTHIA8::Set(``4:m0 = 1.43");
\item PYTHIA8::Set(``5:m0 = 4.30");
\end{itemize}
\section*{References}

\bibliography{mybibfile}

\end{document}